# Enhanced mechanical properties and microstructural stability of ultrafine-grained biodegradable Zn-Li-Mn-Mg-Cu alloys produced by rapid solidification and high-pressure torsion

Wiktor Bednarczyk[a*], Maria Wątroba[b], Grzegorz Cieślak[a], Marta Ciemiorek[a], Kamila Hamułka[b], Claudia Schreiner[c], Renato Figi[c], Marianna Marciszko-Wiąckowska[d], Grzegorz Cios[d], Jakob Schwiedrzik[b], Johann Michler[b], Nong Gao[e], Małgorzata Lewandowska[a], Terence G. Langdon[e]

[a] Warsaw University of Technology, Faculty of Materials Science and Engineering, Wołoska 141, 02-507 Warsaw, Poland
[b] Empa, Swiss Federal Laboratories for Materials Science and Technology, Laboratory of Mechanics of Materials and Nanostructures, Feuerwerkerstrasse 39, CH-3602 Thun, Switzerland
[c] Empa, Swiss Federal Laboratories for Materials Science and Technology, Laboratory for Advanced Analytical Technologies, Überlandstrasse 129, 8600 Dübendorf, Switzerland
[d] Academic Centre for Materials and Nanotechnology, AGH University of Krakow, Al. Mickiewicza 30, 30-059 Krakow, Poland
[e] Materials Research Group, Department of Mechanical Engineering, University of Southampton, Southampton SO17 1BJ, UK
*Corresponding author. E-mail: wiktor.bednarczyk@pw.edu.pl

## Abstract

Zinc alloys have emerged as promising candidates for biodegradable materials due to their remarkable biocompatibility and favorable mechanical characteristics. The incorporation of alloying elements plays an essential role in advancing the tensile strength of Zn alloys. Nevertheless, achieving uniform dispersion of these elements poses challenges due to chemical segregation during solidification. In this study, rapid solidification followed by high-pressure torsion was successfully employed to fabricate Zn-Li-Mn-Mg-Cu alloys characterized by ultrafine-grained microstructures with evenly distributed nanometric intermetallic phases. A comprehensive examination, including phase composition, microstructural evolution, tensile properties and deformation mechanisms, was conducted. The impact of varying annealing temperatures on microstructural stability was systematically examined. The combined implementation of rapid solidification and high-pressure torsion yielded alloys with an average grain size below 360 nm, thereby demonstrating exceptional mechanical properties including yield stress (YS), ultimate tensile strength (UTS), and elongation to failure ($E_f$) equal to at least 325±6 MPa, 350±8 MPa and 40±11 %, respectively. Heat treatment notably augmented the mechanical properties, resulting in a YS = 440±11 MPa and UTS = 491±6 MPa, while preserving plasticity ($E_f$ = 23±4 %) in the Zn-0.33Li-0.27Mn-0.14Mg-0.1Cu alloy. Nanoindentation strain rate jump tests identified thermally activated mechanisms and grain boundary sliding as dominant deformation mechanisms.

## Keywords

# 1. Introduction

For the last two decades significant efforts have been devoted to creating novel biodegradable Zn alloys suitable for resorbable cardiovascular stents and bone-fixing objects [1]. Among the various alloys investigated, Zn alloys offer crucial advantages over both permanent biomaterials and other resorbable materials. Firstly, a bioresorbable implant provides necessary tissue support for a specific period, typically ranging from 6 to 18 months. After this time, the implant undergoes resorption, enabling healthy tissue to heal and function normally again. This feature reduces post-surgery trauma by minimizing the required operations [2]. Secondly, the corrosion process of Zn alloys occurs at an appropriate rate in the human body and does not involve hydrogen gas evolution or harmful corrosion byproduct formation. Other alloys, such as Mg and Fe, have been considered for this application. However, Mg alloys corrode too rapidly and release harmful hydrogen, especially in aqueous environments [3], while Fe alloys exhibit significantly slower corrosion rates [4–7]. The Zn-Li-based alloys appear as the most promising among various Zn alloys due to their extraordinary mechanical properties, good plasticity and proven biocompatibility [8,9].

The preferred and highly effective method for fabricating Zn alloys typically involves a sequence of casting, hot processing, followed by cold processing and subsequent heat treatment [10–13]. This particular approach permits the optimization of both microstructure and mechanical properties. In their as-cast state, Zn alloys tend to be extremely brittle. Therefore, hot processing is necessary to refine the microstructure, enhance the mechanical properties and improve the formability. Cold processing is then employed to further increase plasticity beyond the required level, leveraging the grain refinement effect. However, it should be noted that extensive grain refinement can lead to a reduction in strength in Zn alloys due to the activation of grain boundary sliding (GBS) [14–16]. A controlled grain growth approach addresses this issue by preventing natural aging while enhancing strength [10,17–19].

Eutectic second-phase precipitates play a crucial role in maintaining high mechanical properties by effectively impeding GBS in fine-grained Zn-Mg and Zn-Li alloys [15,20,21]. In both alloys the eutectic mixture is vital in refining the grain size and providing second-phase strengthening. Additionally, the precipitates within the Zn grains act as obstacles, restricting the movement of dislocations and thus further contributing to the overall mechanical strength of the alloys. Based on a literature review, the Zn-Li and Zn-Mg binary systems emerge as the most promising starting points for further modification [8,20]. Notably, the primary focus lies in understanding the impact of small additions of a third element (typically up to 0.4 wt. %) on the overall performance of the alloy. Biocompatible elements can be classified into three main groups based on their major strengthening effects: solid solution strengthening, which is represented by elements such as Ag and Cu; precipitation and second-phase strengthening, including elements such as Li, Mg, Mn; and dispersion strengthening through intermetallic particles, represented by elements such as Ca, Sr, Ti, and Zr [22]. It is worth noting that these elements can each contribute to various strengthening effects. There are literature reports showing remarkable strength in certain complex Zn alloys, such as Zn-0.8Li-0.4Mg (ultimate tensile strength (UTS) equal to 646 MPa and elongation to failure ($E_f$) equal to 3%) and Zn-0.8Li-0.8Mn (UTS = 514 MPa, $E_f$ = 103%) [8]. Therefore, there is a potential for further enhancement by harnessing the mechanical properties of ternary Zn-Li-Mn alloys through additional solid solution strengthening and another source of precipitation strengthening. Cu and Mg exhibit the highest strengthening abilities among the other biocompatible elements.

The addition of a small amount of Mg has been shown to significantly enhance the strength of Zn alloys. For example, in the as-cast state, the addition of 0.1 wt.% of Mg increased the UTS of Zn from 30 to 80 MPa [23]. Even more impressive mechanical properties were achieved in severely cold-drawn Zn-0.08Mg alloy which displayed a UTS exceeding 450 MPa with an $E_f$ below 4% [24]. The Mg addition effectively strengthens Zn alloys by precipitation strengthening of nanometric $Mg_2Zn_{11}$ particles [25]. A Cu addition is also an effective method for strengthening Zn alloys. For a Zn-0.05Mg alloy, the addition of 0.5% Cu increased the UTS from approximately 220 to around 310 MPa [26]. However, it should be noted that a large amount of Cu addition can lead to undesirable phase boundary sliding in ultrafine-grained Zn alloys [27]. Nevertheless, maintaining the Cu addition below the solubility limit in Zn at room temperature (RT) should induce solid solution strengthening without the formation of $CuZn_5$ precipitates [28].

During the solidification of liquid alloys, crystallization takes place. Usually, casting provides a specific morphology-named cast structure. The most characteristic features are a dendritic structure and chemical segregation. To eliminate macroscopic chemical inhomogeneity, annealing or hot deformation is commonly employed. However, these applications are limited for use in Zn alloys. Annealing is effective for Zn-Cu and Zn-Ag alloys due to the substantial increase in solubility with rising temperature. Unfortunately, supersaturated alloys do not exhibit high strength, even after cold deformation, while hot deformation induces precipitation of relatively coarse particles [29]. For hypoeutectic or congruent phase-containing alloys, annealing is not as effective. In these alloys, the changes in solubility are not significant so that the second phases remain unaffected. Additionally, for the Zn-3Ag-0.5Mg alloy there was a rearrangement of the eutectic regions into single grains after die-casting, annealing at 350 °C for 4 hours and subsequent water quenching. Unfortunately, these changes led to mechanical properties that were worse than in the as-cast material [10,30].

There are extensive reports of the challenges involved in achieving a uniform distribution of second phases in the Zn matrix. In all instances, coarse grains or elongated bands of second phases are observed regardless of whether hot or cold processing techniques are used [31,32]. Notably, even subjecting a Zn-3Ag-0.5Mg alloy to high-pressure torsion for 5 turns (total strain ~100) under an applied pressure of 6 GPa failed to crush the Mg-rich particles [30]. This phenomenon is attributed to the significant difference in hardness between the soft matrix and the hard phases. During plastic forming, the Zn-based matrix deforms under stresses lower than the minimum required to crush the second-phase particles. The presence of inhomogeneously distributed second phases weakens the grain boundary strengthening effect, leading to a relatively high contribution from GBS and a reduction in the mechanical properties. Ideally, the optimal microstructure should consist of fine, uniformly dispersed second-phase precipitates located at grain boundaries. The chemical composition and crystallization conditions influence the final size of second-phase particles after casting. An effective method for achieving a finer microstructure is by increasing the cooling rate through a melt-spinning casting approach, also known as rapid solidification. This technique allows for cooling rates in the $10^4 \div 10^6$ K s$^{-1}$ range, leading to non-equilibrium crystallization and the formation of an overcooled, metastable microstructure. As a result, rapid solidification of alloys brings about significant changes in their structures, including enhanced compositional uniformity, microstructural refinement, metastable phases and very fine dispersions of second-phase particles.

Rapid solidification has been utilized on several occasions to fabricate Zn alloys, with most studies focusing on the effects of rapid solidification and subsequent annealing on the microstructure and properties of melt-spun ribbons. This research has particularly centered around eutectic Zn-Mg alloys [32], peritectic Zn-Ag alloys [33] and eutectoid Zn-Al [34] due to the different types of reactions during solidification. In each case, rapid solidification led to the formation of non-equilibrium, fine-grained microstructures with well-dispersed second-phase particles. An impressive example of effective grain refinement and homogenous distribution of second-phase precipitates was reported for Zn-Mg alloys [32]. The hot-compacting of melt-spun ribbons resulted in a significantly smaller grain size than the hot extrusion of the as-cast material. Moreover, a uniform distribution of $Mg_2Zn_{11}$ precipitates was achieved after hot processing. The grain refinement obtained through rapid solidification was reduced to approximately 2 μm due to the relatively high temperature employed during the process.

The most effective approach for producing bulk materials from melt-spun ribbons involves a cold-deformation technique. However, only severe plastic deformation (SPD) methods can generate the necessary hydrostatic pressure for successful room temperature consolidation [35]. Among the various SPD techniques, high-pressure torsion (HPT) stands out as the most promising tool since it offers both high pressure and low temperature for fast processing. HPT is frequently employed to create ultrafine-grained and nanocrystalline materials due to its ability to apply significant strain values relatively quickly [36]. This makes HPT a versatile and advantageous method for achieving the desired material properties from melt-spun ribbons.

The present study was initiated to produce and analyze bulk high-strength Zn-Li-Mn-X alloys using a combination of melt spinning and HPT where this is a previously never-explored procedure for Zn alloys. Thus, this research represents the first attempt to utilize both methods in tandem with these materials. Previous investigations have established the excellent performance of conventionally-processed Zn-Li-Mn alloys [8] but this study takes the analysis further by examining the mechanical and thermal stability characteristics of ultrafine-grained microstructures. Additionally, this research explores the impact of minor additions of Mg and Cu on the mechanical properties of the alloys, which will provide precipitation strengthening and solid solution strengthening, respectively, without forming an extensive volume of intermetallic particles.

## 2. Materials and Methods
### 2.1. Materials preparation

The alloys investigated in this study were Zn-0.33Li-0.39Mn (S1) and Zn-0.33Li-0.27Mn-0.14Mg-0.1Cu (S2). To prepare the alloys, pure Zn (99.995 wt.%), Li (99.95 wt.%), Mn (99.9 wt.%), Mg (99.95 wt.%), and high-purity Cu-40Zn brass (<0.007 wt.% of impurities) were used. The alloys were produced through a melting process in a graphite crucible at a temperature of 650 °C. The molten alloy was then cast into a cylindrical steel mold. After casting, the alloys were subjected to an annealing treatment at 350 °C for 4 hours. Subsequently, the annealed alloys were rapidly quenched in water to achieve a uniform chemical distribution. The chemical composition of the investigated materials was determined using Agilent 5110 inductively coupled plasma optical emission spectroscopy (ICP-OES) where 0.5 grams of the samples were completely dissolved in 5 mL hydrochloric acid 37% p.a. (Merck) in a beaker under gentle heating. The samples were then filled with ultrapure water (Millipore 18.2 MΩcm) in a 500 mL volumetric flask and the quantitative determination of the elements sought was carried out by ICP-OES. Table 1 presents the measured chemical composition of the alloys. The phase composition of the investigated annealed, coarse-grained alloys was measured using a Panalytical Empyrean CuKα X-ray diffractometer (XRD). Data were collected with a scanning rate of 0.4°/min and a step size of 0.02°.

The melt-spinning technique was employed to achieve rapid solidification of the alloys. The process was conducted under an argon atmosphere with a copper wheel speed set at 30 m/s. This led to the formation of thin ribbons from the molten alloys (denoted as RS). Samples for the HPT process were prepared using two methods. Firstly, samples were directly cut from conventionally homogenized cast rods (denoted as CC and HPT-CC pre- and post-HPT processing, respectively). Alternatively, samples were cold compacted from the melt-spun ribbons under a pressure of 100 MPa. The HPT process was performed at room temperature under a pressure of 6 GPa with a rotation speed of 1 revolution per minute. The HPT was conducted for 15 turns to obtain good mechanical integrity and microstructure homogeneity (samples denoted as HPT-RS). Additionally, samples after HPT were annealed at 140°C, 165°C, 190°C or 220°C (herein denoted as HPT-CC-HT or HPT-RS-HT) for 10 minutes with the temperatures measured using a thermocouple placed next to the samples.

*Table 1 Chemical composition of investigated alloys measured by ICP-OES (all values ±<0.01 wt. %)*

| Material | Zn (wt./at. %) | Li (wt./at. %) | Mn (wt./at. %) | Mg (wt./at. %) | Cu (wt./at. %) |
|---|---|---|---|---|---|
| S1 alloy | bal. | 0.33/3.0 | 0.39/0.45 | - | - |
| S2 alloy | bal. | 0.33/3.0 | 0.27/0.31 | 0.14/0.37 | 0.1/0.1 |

### 2.2. Samples preparation

Cross-sectional samples were prepared for microstructural observations using the standard metallographic preparation method. The preparation involved a series of steps using abrasive papers and then water-free diamond suspensions with 3 µm and 1 µm particle size for polishing. The final step in the preparation process was electrolytic polishing which was carried out for 10 – 20 seconds in a solution consisting of 6% perchloric acid in ethanol at a temperature of -35 °C and a voltage of 30 V. The same procedure was utilized to obtain transparent thin-foils for transmission electron microscopy using a twin-jet electropolishing system. It was necessary to perform the polishing immediately before observation due to the high oxidation tendency of the samples. Tensile samples were cut from HPT disks using wire electro-discharge machining (WEDM), using the procedure illustrated schematically in Fig. 1.

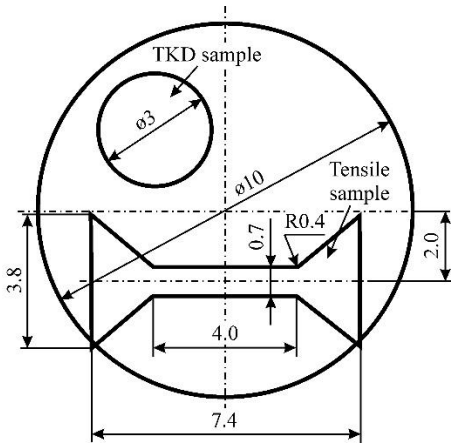

*Fig. 1 Schematic illustration of the location of the tensile testing specimen and TKD disk on the HPT-processed sample.*

### 2.3. Microstructure characterization

Microstructure imaging was performed using an Hitachi SU-70 field-emission scanning electron microscope (SEM). The qualitative elements distribution was measured using a Thermo Scientific UltraDry energy dispersive X-ray spectroscopy (EDS) detector. Additionally, off-axis transmission Kikuchi diffraction (TKD) measurements were performed using a Bruker Quantax eFlashHD electron backscattered diffraction (EBSD) detector installed in the SEM. TKD maps were collected using the following parameters: acceleration voltage 30 kV, step size 28 nm, sample inclination angle 0°. Pattern indexing was performed using Oxford Instruments AztecCrystal MapSweeper software by means of dynamic template matching [37]. All patterns were indexed as Zn phase as the EBSD differentiation between Zn and LiZn$_4$ is complex and is under ongoing investigation. Grains were defined as a set of at least six points surrounded by an uninterrupted grain boundary with a misorientation of at least 15°. The average grain size of the HPT-CC and HPT-RS samples was calculated based on the area fraction distribution using ATEX software [38,39]. The distinction between low-angle grain boundaries (LAGB) marked in grey and high-angle grain boundaries (HAGB) marked in black was established by considering the disorientation angle between grains, with LAGB (gray lines) defined as disorientation angles ranging from 3° to 15° and HAGB (black lines) defined as disorientation angles spanning from 15° to 95°. An average grain size of HPT-CC-HT and HPT-RS-HT samples was approximately estimated using an intercept procedure.

### 2.4. Mechanical testing

The mechanical properties of the samples were investigated through a series of tests, including hardness measurements, nanoindentation mapping, nanoindentation strain rate jump tests and uniaxial tensile testing. Vickers hardness tests were performed using a ZwickRoell DuraScan G2 hardness tester. The tests were conducted at a distance of 1.5 mm from the edge of the disk under a force of 0.98 N (Hv0.1). The hardness distributions across the cross-section of the disks were measured using a ZwickRoell ZHN Nanoindenter with the measurements performed in the load-control mode under a load of 50 mN. Strain rate sensitivity (SRS) measurements were carried out using the nanoindentation strain rate jump (SRJ) method [40,41]. The experiments used a displacement-controlled Alemnis Scanning Nanoindenter operating in continuous stiffness measurement mode (CSM) at an oscillation frequency of 10 Hz and an amplitude of 10 nm. Each test was performed at the following strain rates: $5 \times 10^{-2}$ s$^{-1}$ (base strain rate), $1 \times 10^{-1}$ s$^{-1}$, $5 \times 10^{-3}$ s$^{-1}$ and $5 \times 10^{-4}$ s$^{-1}$. The depth of each segment was adjusted to ensure stable results. Each test was repeated six times for each sample and at least four similar curves were selected for analysis. Fig. 2 presents an example of the experimental curves (Fig. 2a) and the calculated reduced Young's modulus and hardness plots with respect to contact depth (Fig. 2b,c). A constant value of Young's modulus at all strain rates and constant hardness at the strain rate of $5 \times 10^{-2}$ s$^{-1}$ confirms the validity of the measurements. All indentation tests were performed on the same samples for microstructure analysis; therefore, the mechanical properties were measured on the cross-sections of the HPT-processed disks. The mechanical properties of the samples were further assessed through uniaxial tensile testing of dog-bone-shaped miniature specimens with gauge dimensions of $0.7 \times 0.6 \times 4.0$ mm at a strain rate of $10^{-3}$ s$^{-1}$ and at room temperature using a Zwick Roell Z005 universal tensile machine.

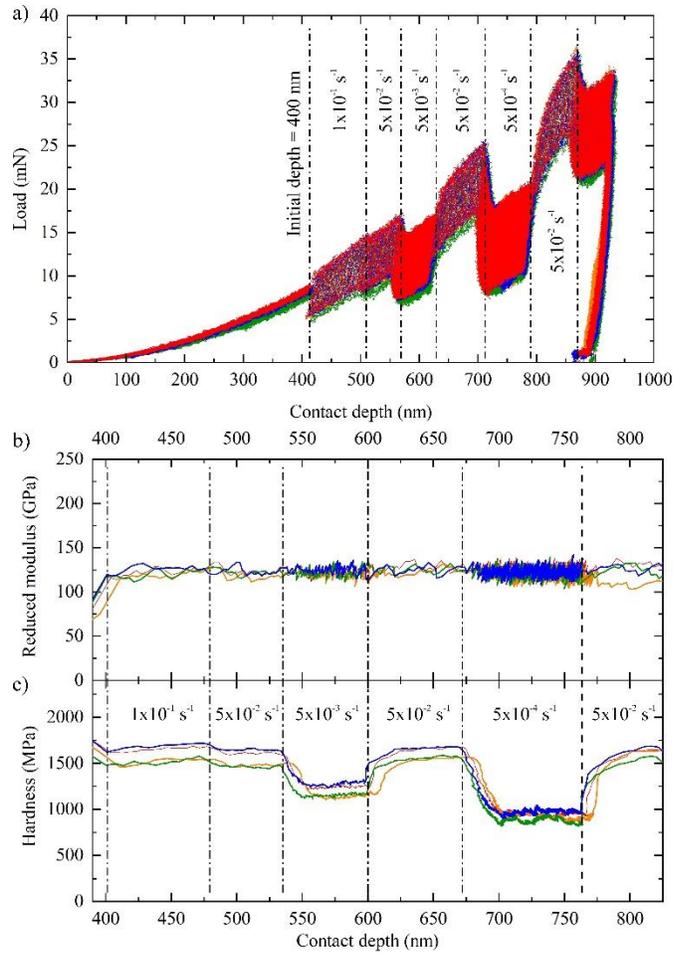

*Fig. 2 Example of SRJ test curves with highlighted procedure steps (a), and calculated reduced Young's modulus (b) and hardness (c) in respect to the depth (S1 HPT-RS-HT sample).*

## 3. Results

### 3.1. Characterization of the as-cast state

Fig. 3 presents X-ray patterns of the annealed S1 and S2 alloys. Both alloys are composed of a Zn matrix (hexagonal close-packed - HCP, $c/a$ = 1.856), αLiZn$_4$ phase where the crystal structure was different from the literature and was corrected according to the XRD results (Space group *Amm2* - superstructure derived from *Cmcm* [42,43], $a$ = 4.43160 Å, $b$ = 19.39000 Å, $c$ = 4.81260 Å, α = β = γ = 90°), minor βLiZn$_4$ (HCP, $c/a$ = 1.598), MnZn$_{13}$ (space group 12, $a$ = 13.483 Å, $b$ = 7.662 Å, $c$ = 5.134 Å, β= 127.78º) and Li$_2$O$_2$ (space group 174, $a$ = 6.305 Å, $c$ = 7.710 Å) phases. The MnZn$_{13}$ phase was barely detected in both alloys while the Mg- and Cu-containing phases were not detected in the S2 alloy because their volume fraction was below the detection limit.

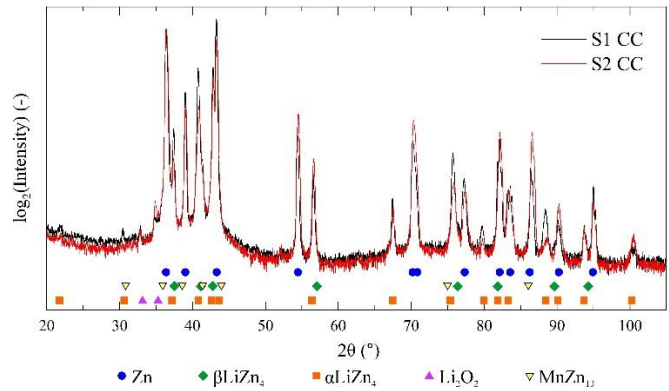

*Fig. 3 X-ray diffractograms of annealed, coarse-grained S1 and S2 alloys. $Cu_{K\alpha}$ radiation.*

Fig. 4 illustrates the microstructures of the initial as-cast annealed and rapidly solidified materials. In the S1 CC alloy (Fig. 4a,b), the microstructure consists of large Zn grains with plate-like nanometric $LiZn_4$ precipitates and a visible substructure in the $LiZn_4$ grains (Fig. 4b). Surprisingly, EDS analysis presented in Fig. 5a, indicates a subtle enrichment of $LiZn_4$ in Mn. However, it may result from a Zn depletion in this region thereby giving an EDS analysis artifact. The hardness of the S1 alloy equals 102 Hv0.1. Rapid solidification reduced the grain size and changed the grain shape to elongated in the cooling direction. Moreover, rapid solidification significantly altered the phase distribution. The $LiZn_4$ phase precipitated as elongated thin plates within the Zn grains rather than forming distinct grains (Fig. 4e,f). Rapid solidification also caused a significantly greater homogenous distribution of Mn in the microstructure.

For the S2 CC alloy (Fig. 4c,d), the addition of small amounts of Mg and Cu had a notable effect on the microstructure and the hardness. The grain size was significantly refined to below 10 µm and there were small Mg-rich grains, indicated by red arrows, confirmed by EDS analysis presented in Fig. 5c and nanometric precipitates (indicated by red arrows in Fig. 4d). Apart from the grain size and Mg-rich particles, the microstructure was similar to that of the S1 alloy. Small Mg and Cu additions significantly increased the hardness to 121 Hv0.1. The S2 RS alloy demonstrated significant changes in the microstructure compared to the annealed state (Fig. 4g,h). Equiaxed $LiZn_4$ grains were present between relatively small Zn grains and as nanometric precipitate needles within the Zn grains. Importantly, rapid solidification hindered the precipitation and growth of Mg-rich particles which are no longer invisible (Fig. 5d).

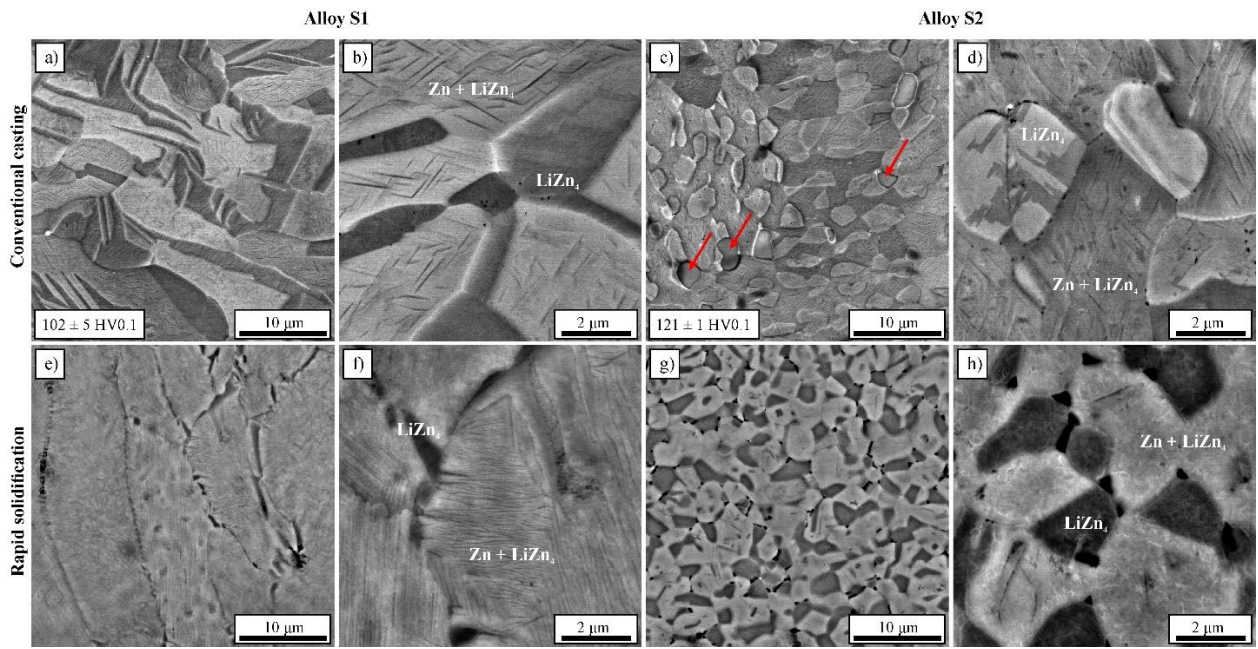

*Fig. 4 SEM-BSE images of annealed (a-d) and rapidly-solidified (e-h) S1 (a,b,e,f) and S2 (c,d,g,h) alloys. Red arrows indicate Mg-rich phases.*

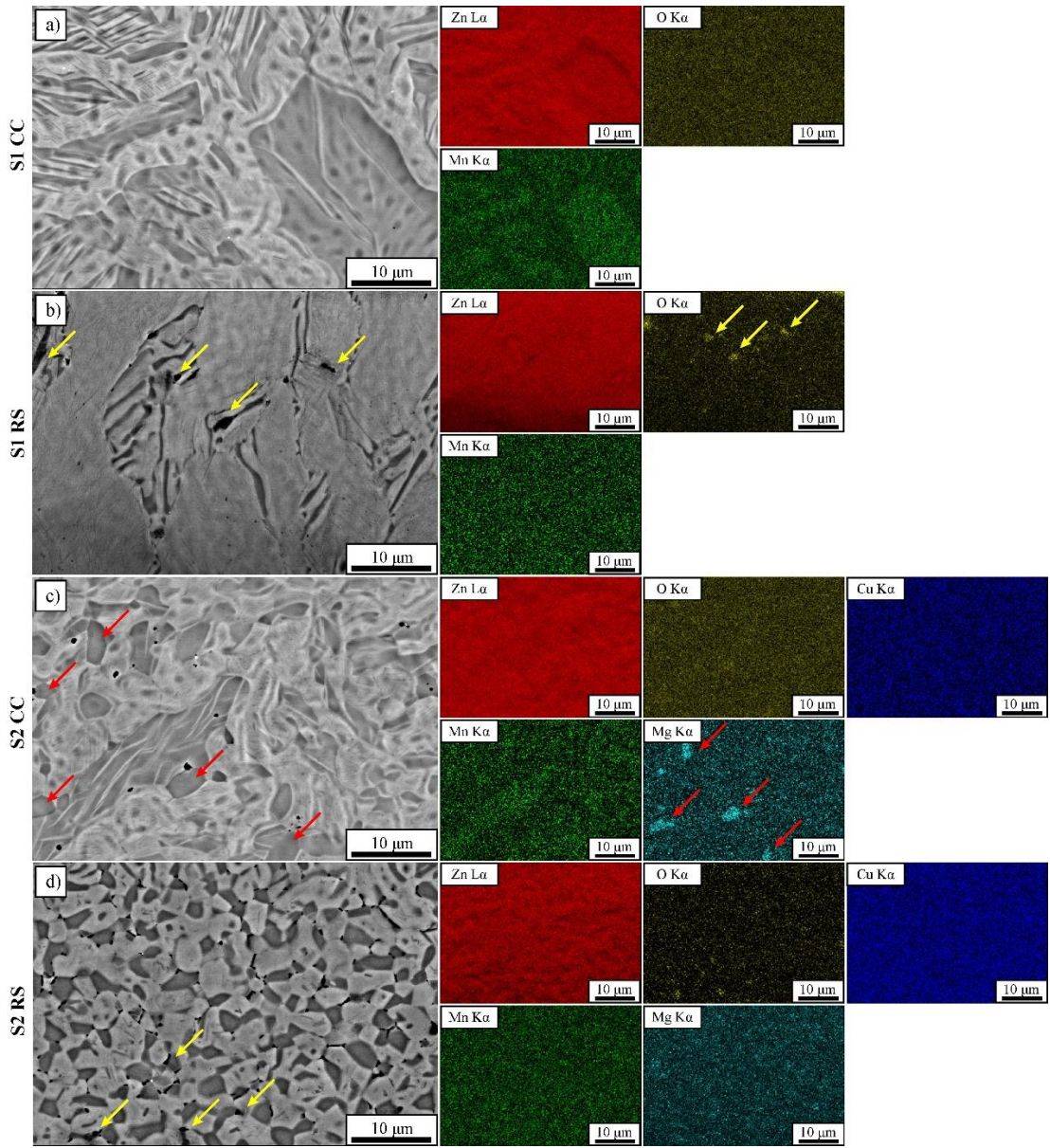

*Fig. 5 EDS analysis of S1 CC alloy (a), S1 RS alloy (b), S2 CC alloy (c), S2 RS alloy (d). Yellow and red arrows indicate oxides and Mg-rich particles, respectively.*

### 3.2. Microstructure characterization after HPT

Severe plastic deformation through HPT led to significant grain refinement in the as-cast and rapidly solidified states. The microstructure after HPT processing was analyzed using SEM-BSE (Fig. 6) and TKD analyses (Fig. 7). Fig. 6a-d illustrates the impact of HPT on the microstructure of the HPT-CC samples. After 15 turns, a uniform microstructure consisting of small equiaxed recrystallized grains, slightly elongated in the shear direction, was observed in both alloys. However, distinguishing between the Zn and $LiZn_4$ phases was not possible using either SEM-BSE imaging or TKD. Nonetheless, coarse grains of Mg-rich phases remained present in the microstructure of the S2 HPT-CC alloy even after 15 turns (Fig. 6d). This significant grain refinement resulted in increased hardness for both alloys. After 15 turns, there were hardness values of 116±2 Hv0.1 and 129±1 Hv0.1 for the S1 and S2 alloys, respectively. Analysis of the S1 HPT-RS alloy revealed visible microstructural changes without notable changes in grain size. Fig. 6e provides an overview of the microstructure of the S1 HPT-RS alloy showing the presence of oxide layers (indicated by yellow arrows) originating from the oxidation of the RS ribbon surfaces. A more pronounced effect of rapid solidification was observed in the S2 HPT-RS alloy (Fig. 6g,h), where oxide layers may be present and the Mg-rich grains were refined below the observation limit of SEM-BSE. Rapid solidification conducted prior to HPT effectively enhanced the hardness in both alloys to 121±1 Hv0.1 and 141±1 Hv0.1 for the S1 and S2 alloys, respectively. TKD analysis (Fig. 7) revealed the influence of both processing and chemical composition on the grain size and microtexture observed on the shearing plane. Except for the S2 HPT-RS alloy, all samples exhibited a measured grain size of approximately 330-370 nm. The S2 HPT-RS alloy showed a smaller grain size measured at 250 nm. The variation in grain shape observed in TKD and SEM-BSE imaging is attributed to the observation planes lying perpendicular in these examinations. Based on the TKD analysis inverse pole figures (IPF) were calculated and presented in Fig. 7e-h. These IPF images show a similar $\{0001\}\langle11\bar{2}0\rangle$ texture in all HPT-processed samples. However, the HPT-CC samples exhibit a higher texture intensity than the HPT-RS samples.

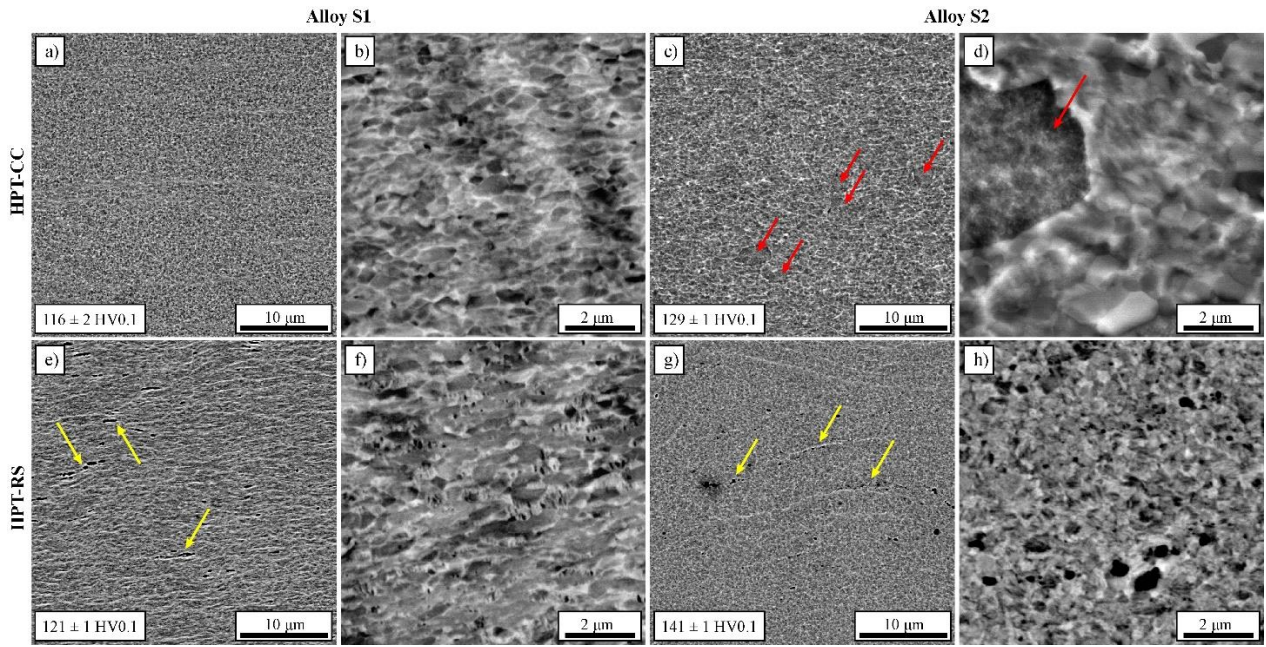

*Fig. 6 SEM-BSE images of HPT-CC (a-d) and HPT-RS (e-h) of S1 (a,b,e,f) and S2 (c,d,g,h) alloys. Yellow arrows indicate oxides in HPT-RS samples while red arrows indicate Mg-rich particles in S2 HPT-CC sample.*

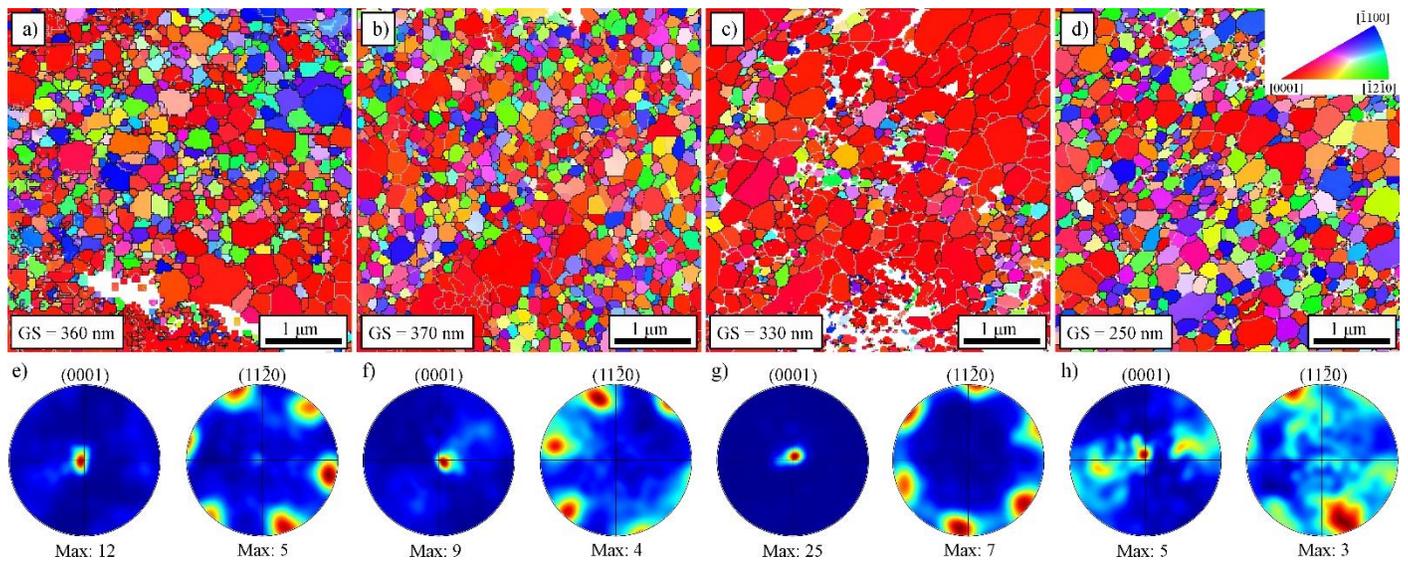

*Fig. 7 Top row: TKD maps of HPT-processed alloy: S1 HPT-CC (a), S1 HPT-RS (b), S2 HPT-CC (c), S2 HPT-RS (d). Bottom row: Corresponding inverse pole figures for each TKD map. Gray and black lines indicate LAGBs and HAGBs, respectively.*

A non-uniform microstructure is generated across the diameter of the disk during HPT processing. Thus, the distributions of microhardness across cross-sections were measured on HPT-processed samples. Fig. 8a,b illustrates the microhardness of the S1 alloy. Increasing the applied strain increased the microhardness from approximately 0.8 GPa to around 1.0 GPa in both the HPT-CC and HPT-RS states. Furthermore, the changes in microhardness were more regular in the HPT-CC samples compared to the HPT-RS sample. Fig. 8c,d demonstrates a significant impact of both the chemical composition and processing route on the microhardness distribution in the S2 alloy after HPT. In the HPT-CC state, the increased strain decreased the microhardness from approximately 1.3 GPa to around 1.1 GPa. A similar trend was observed in the HPT-RS state. Surprisingly, recrystallization led to an overall decrease in microhardness. In the center of the disk the microhardness was approximately 1.0 GPa whereas at the edge of the disk it was around 0.9 GPa.

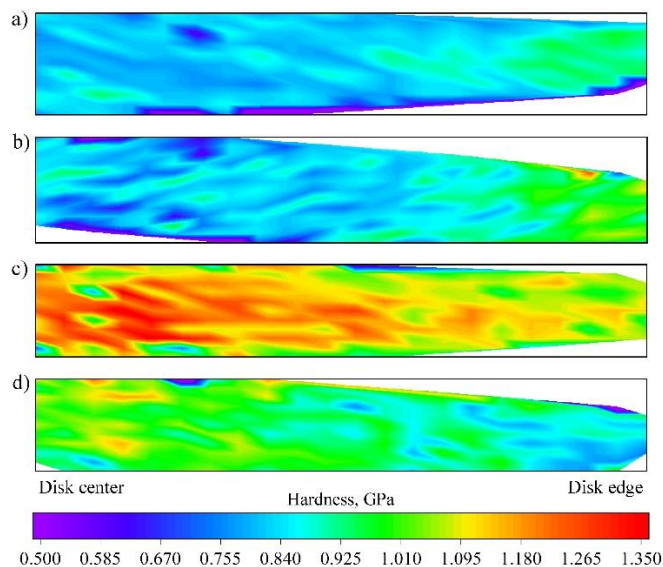

*Fig. 8 Nanoindentation hardness distribution on cross-sections of HPT-processed disks. S1 HPT-CC (a), S1 HPT-RS (b), S2 HPT-CC (c) and S2 HPT-RS (d) samples.*

### 3.3. Effect of post-deformation annealing

Fig. 9 illustrates the impact of heat treatment on the microstructure of the HPT-RS samples. In the S1 alloy, only slightly visible changes in the microstructure were observed up to a temperature of 165 °C. However, as the temperature increased, noticeable grain growth up to an estimated approximately 0.8 µm and 1.5 µm was observed after annealing at 190 °C and 220 °C, respectively. Fig. 9c displays a heterogeneous microstructure consisting of coarse recrystallized grains and fine-grained regions in the S1 alloy. Increasing the annealing temperature led to homogeneous grain growth (Fig. 9d). However, this observed grain growth had no significant effect on the hardness which remained around 114±2 Hv0.1. This indicates the impact of grain refinement softening, where grain growth is expected to produce an increased hardness coupled with a decrease in solid solution strengthening due to the redistribution of the alloying elements.

Fig. 9e-h depicts the effect of annealing on the HPT-RS S2 alloy. An increase in annealing temperature in this alloy led to gradual microstructure and grain size changes. Grain growth was initiated at ~140 °C and became evident at 165 °C, producing grains with a size estimated at approximately 1 µm. Further temperature increase led to grain growth with grains exceeding 2 µm at 220 °C. Surprisingly, this gradual grain growth failed to produce proportional changes in hardness. No hardness changes were observed up to 165 °C but a notable drop in hardness to 128 Hv0.1 was observed after annealing at 190 °C. Subsequent temperature increases produced no additional changes in the measured hardness. A heat treatment at 190 °C for 10 minutes was chosen

for further analysis of the microstructural stability of HPT-CC samples and mechanical investigation. Fig. 10 presents a microstructure comparison of the HPT-CC-HT and HPT-RS-HT samples. Heat treatment caused a significantly more pronounced grain growth to ~2.0 µm and ~2.5 µm in the S1 and S2 HPT-CC-HT samples, respectively, than in HPT-RS-HT samples.

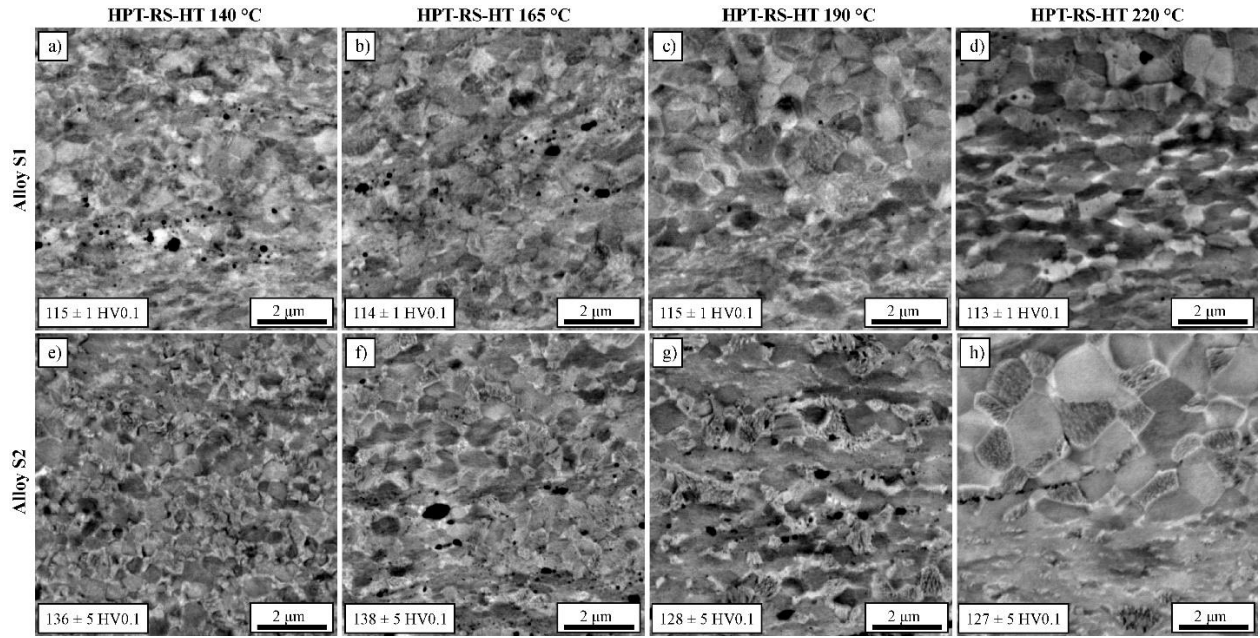

*Fig. 9 SEM-BSE images of HPT-processed and heat-treated (HPT-RS-HT) samples of S1 (a-d) and S2 (e-h) alloys at 140 °C (a,e), 165 °C (b,f), 190 °C (c,g), and 220 °C (d,h) for 10 minutes.*

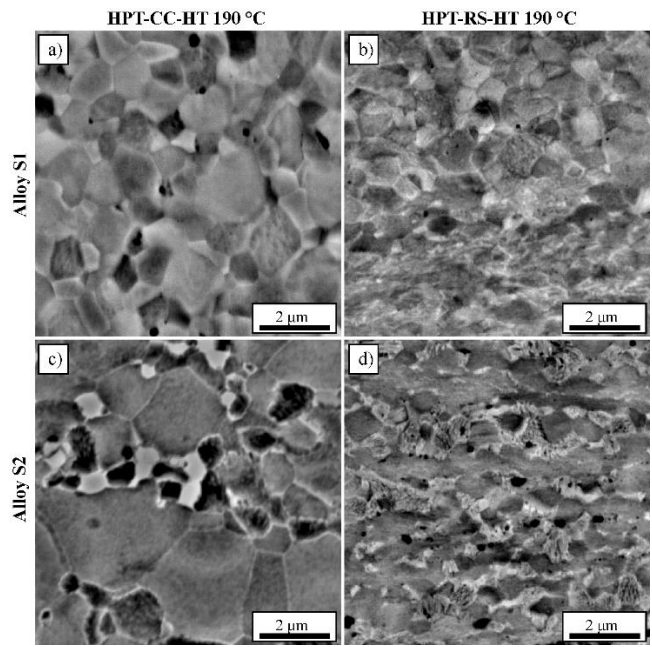

*Fig. 10 Comparison of microstructure stability after heat treatment at 190 °C for 10 minutes in HPT-CC-HT (a) and HPT-RS-HT (b) S1 alloy and HPT-CC-HT (c) and HPT-RS-HT (d) S2 alloy*

### 3.4. Mechanical characterization

Fig. 11 illustrates the results of the tensile tests conducted on HPT-processed and heat-treated at 190 °C alloys in the HPT-CC and HPT-RS states. The choice of the heat treatment was based on the observed microstructure and hardness changes in both alloys in the HPT-RS samples. In the HPT-CC state, the S1 alloy exhibited relatively low mechanical properties and good plasticity, with yield stress (YS), ultimate tensile strength (UTS) and elongation to failure ($E_f$) of approximately 265±4 MPa, 283±3 MPa and 92±2%, respectively (Fig. 11a). Rapid solidification significantly increased the strength while slightly reducing the plasticity, leading to YS = 324±6 MPa, UTS = 350±8 MPa and $E_f$ = 40±6%. The heat treatment performed on the HPT-CC samples caused significant changes in their mechanical properties. A substantial increase of 22% in YS and 31% in UTS was observed, giving YS = 322±4 MPa and UTS = 370±3 MPa (Fig. 11c). The HPT-RS samples showed a less pronounced effect with no change in YS and a 4% change in UTS (363±2 MPa).

The S2 alloy, containing Mg and Cu additions, exhibited higher mechanical properties than the S1 alloy (Fig. 11b). In the HPT-CC state, the YS and UTS were approximately 323±6 MPa and 350±11 MPa, respectively, whereas in the HPT-RS samples they were 363±3 MPa and 382±3 MPa, respectively. Despite the higher mechanical strength, the plasticity of the S2 alloy remained similar to that of the S1 alloy. Furthermore, the heat treatment produced similar changes in the mechanical properties for both S1 and S2 alloys. The S2 HPT-CC-HT alloy demonstrated outstanding mechanical properties, with YS = 440±11 MPa and UTS = 490±5 MPa representing 36% and 40% increases compared to the HPT-CC state, respectively. The changes observed in the S2 HPT-RS-HT alloy were significantly less pronounced, with a 3% increase in YS (375±1 MPa), a 9% increase in UTS (416±4 MPa) and a reduction in elongation to failure to 27% due to the heat treatment.

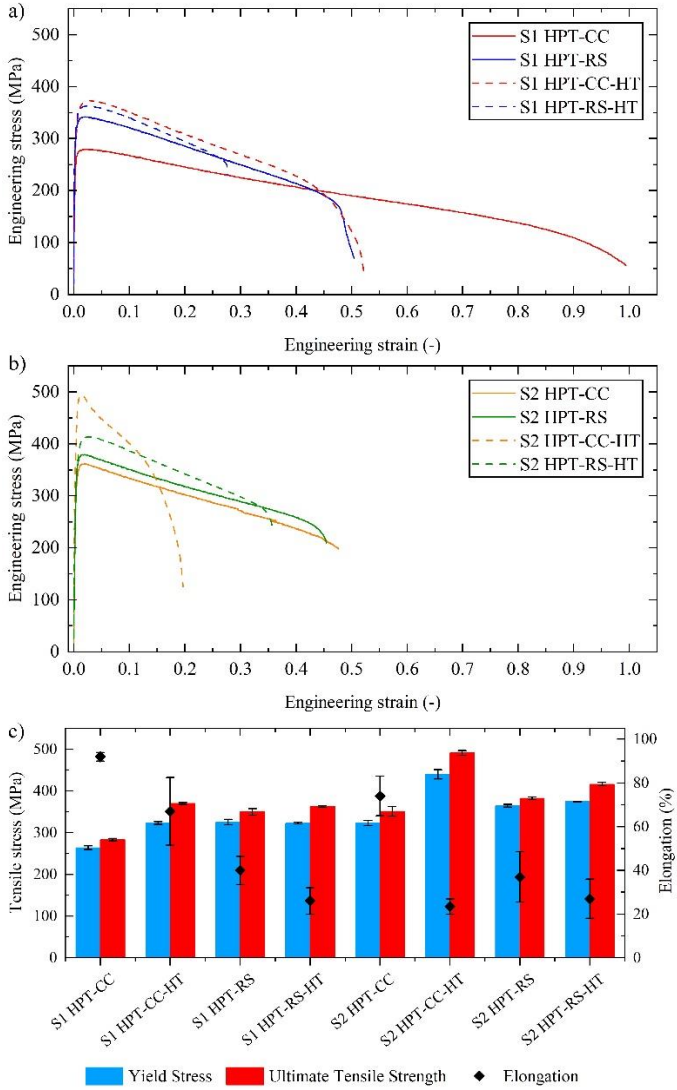

*Fig. 11 Engineering strain–stress curves of S1 (a) and S2 (b) HPT-CC, HPT-RS and HPT-CC/RS-HT alloys. Comparison of mechanical properties of S1 and S2 alloys (c)*

The results presented in Fig. 12 demonstrate the impact of different processing routes and heat treatments on the deformation mechanisms as determined by the SRS parameter calculated by the results from the strain rate jump tests. The strain rate sensitivity *m* was calculated from the slope of the logarithmic hardness vs. strain rate plot:

$$m = \frac{\partial \ln \sigma}{\partial \ln \dot{\varepsilon}} \sim \frac{\partial \ln H}{\partial \ln \dot{\varepsilon}} = \frac{\ln H_2 - \ln H_1}{\ln \dot{\varepsilon}_2 - \ln \dot{\varepsilon}_1} \qquad (1)$$

where *H* is the hardness value and $\dot{\varepsilon}$ is the strain rate at the given segment. The SRS parameter for the S1 HPT-CC alloy is 0.10 at the highest strain rate range and 0.17 at the lower strain rate ranges. By contrast, the S1 HPT-RS exhibits an *m* value that is more than 20% lower than the HPT-CC samples across all strain rate ranges. Additionally, the heat treatment of the HPT-RS samples further reduces the *m* value, with the most significant effect observed at 140 °C where m equals 0.05 and 0.10 at the highest and lower strain rate ranges, respectively. As the annealing temperature increases, the *m* value gradually rises and reaches *m* = 0.08 and approximately 0.12 in the strain rate ranges of $1 \times 10^{-1} - 5 \times 10^{-2}$ s$^{-1}$ and $5 \times 10^{-2} - 5 \times 10^{-3}$ s$^{-1}$, respectively.

Fig. 12b presents the measured values of the SRS parameter *m* for the S2 alloy. The S2 alloy exhibits higher *m* values in all states compared to the S1 alloy. The highest SRS is observed in the HPT-processed samples with *m* values reaching 0.20. Furthermore, the HPT-RS state induces a significantly higher SRS at the highest strain rate range, with *m* = 0.14 compared to *m* = 0.10 in S1. Heat treatment of the S2 alloy at 140 °C leads to a substantial reduction in SRS at high strain rates and a less pronounced effect at lower strain rates. An increase in the annealing temperature leads to more pronounced changes in the SRS parameter *m* in the S2 HPT-RS alloy compared to the S1 HPT-RS alloy.

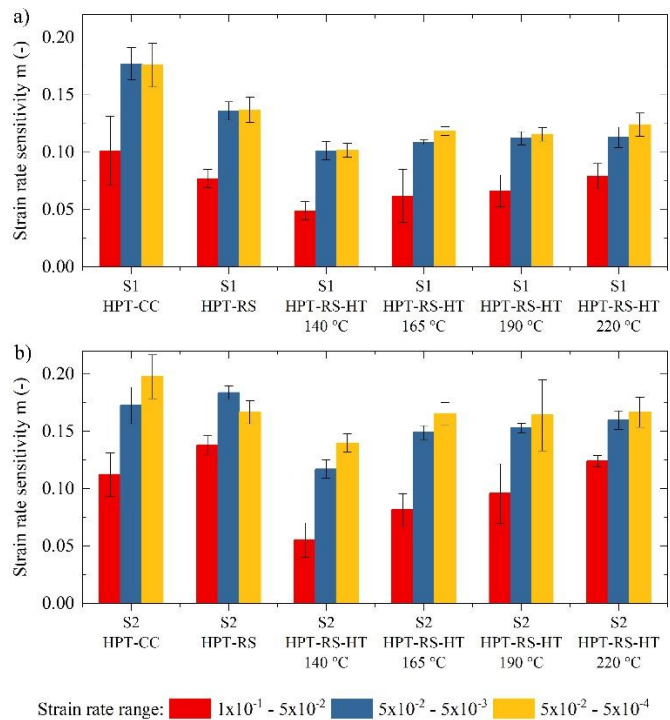

*Fig. 12 Strain rate sensitivity analysis of S1 (a) and S2 (b) alloys based on nanoindentation strain rate jump tests at room temperature.*

## 4. Discussion

### 4.1. The effect of the processing route on microstructure development

The cast and annealed alloys feature the presence of large grains of Zn and eutectic Zn-LiZn$_4$. These observed microstructures correspond well with other reports [44]. Small plates of LiZn$_4$ were observed within the Zn grains which led to changes in the Li solubility in Zn with temperature. Annealing at 350 °C for 4 hours caused a substantial fraction of the alloying elements to dissolve in the Zn matrix leading to its precipitation during cooling. Additionally, twin-like structures were observed in the LiZn$_4$ grains (Fig. 4b and d) suggesting, as expected from the phase diagram, a decomposition of βLiZn$_4$ to αLiZn$_4$ and Zn phases [45,46]. The XRD results confirmed the presence of two phases identified as αLiZn$_4$ and βLiZn$_4$ with theoretical *c/a* ratios of about 1.598 and 1.580, respectively, which is slightly lower than the perfect 1.63 value of *c/a* ratio for HCP metals. This implies good plasticity of LiZn$_4$ via multiple twinning and slip systems which are restricted in Zn [47]. The Zn matrix enrichment in Mn, Mg and Cu caused intermetallic phases precipitation but their detection was challenging. However, these phases cannot be discarded from further analysis based only on the XRD results because their amount may be below the detection limit. Additionally, Mg-rich grains were observed in the S2 CC alloy. Furthermore, Fig. 5a and c illustrate a nuanced enhancement of Mn within the LiZn$_4$ grains. Given the notably limited solubility of Mn in Zn at room temperature, any surplus Mn may conceivably find its way into the LiZn$_4$ phase. Nevertheless, the observed phenomenon could also follow from a depletion of the Zn matrix in these regions, where Li substitution occurs in 20% of Zn atom sites within the LiZn$_4$ grains. This could lead to reduced signal absorption and an elevated Mn count rate during EDS mapping. Notably, this phenomenon was absent in the RS alloys due to the more uniform distribution of the LiZn$_4$ phases.

The main goal of the rapid solidification in this research was to refine all phases except the Zn matrix due to the challenging task of achieving ultra-fine grains in the Zn matrix due to its low melting temperature (~420 °C). Fig. 4e-h illustrates significant changes in the microstructure as a consequence of the rapid crystallization conditions. In the S1 alloy the LiZn$_4$ formed as nanometric thick plates whereas in S2 it primarily appeared as equiaxed grains and nanometric precipitates within the Zn grains. Additionally, rapid solidification led to the refinement of coarse Mg-rich grains which is similar to results in previous experiments [32]. The EDS analysis confirmed its beneficial effect on the uniform distribution of Mg within the microstructure (Fig. 5).

A novel fabrication approach was proposed in this investigation combining rapid solidification with subsequent HPT processing and this produced remarkable grain refinement to approximately 350 nm (Fig. 6). The microstructure obtained using this method demonstrated exceptional uniformity compared to the conventional production of similar alloys [13]. This uniformity was attributed to continuous dynamic recrystallization during the deformation process and the remarkable refinement of Mg-rich particles during melt spinning. The nanometric and evenly distributed Mg-rich particles appear to play a significant role in grain refinement and grain size retention during heat treatment. Several earlier results demonstrated numerous examples of Zn-Mg alloy processing, either hot or cold, where $Mg_2Zn_{11}$ agglomerates were poorly distributed and crushed [31,48,49]. Even after SPD processing, these agglomerates produced bands of slightly refined $Mg_2Zn_{11}$ particles within the Zn matrix [50–52]. Complex Mg-rich grains also remained unchanged even after HPT processing both in the present experiments (Fig. 6d) and elsewhere [30]. Additionally, experiments using powder metallurgy and subsequent hot extrusion exhibited particle bands in Zn-Mg alloys [53].

Fig. 7 demonstrates a practical effect of reducing the size of Mg-rich particles on grain refinement by 25% in the HPT-RS compared to the HPT-CC S2 alloy. The only apparent microstructural difference is the size of Mg-rich particles (see Fig. 6d and h). Moreover, rapid solidification alone cannot be considered a primary refinement factor as in the S1 alloy the grain size remains independent of the casting method. Similar results were presented in a Zn-1.6Mg alloy produced by melt-spinning and hot extrusion [32]. Nevertheless, despite significantly higher Mg content, the measured grains size was about 2 μm. Obviously, the alloys in the present investigation were cold processed by HPT compared to hot extrusion [32]. Nevertheless, HPT processing appears to be the only possibility to compact melt-spun ribbons at room temperature without significant grain growth. Also, a Li addition seems to positively affect uniform grain refinement due to the lower hardness dissimilarities between the $LiZn_4$ and Zn phases than other phases, for example, $Mg_2Zn_{11}$ and $CuZn_5$, in promoting simultaneous deformation of both phases.

Microstructure development can be observed using not only microscopic methods but also mechanical mapping. Fig. 8 illustrates the hardness distributions in the HPT-CC and HPT-RS states of the S1 and S2 alloys. For the S1 samples, an increase in strain leads to a corresponding increase in hardness while the opposite effect is observed in the S2 samples. The strain hardening in the S1 alloy can be attributed to the strain-induced dissolution of $MnZn_{13}$ precipitates and, consequently, grain growth. On the other hand, the higher hardness in S2 results from the addition of Mg and Cu but strain softening occurs due to significant grain refinement at the disk edge. It is important to note that grain refinement in an ultrafine-grained Zn alloy does not follow the Hall-Petch (H-P) relationship. The H-P relationship breaks down at specific grain sizes: approximately 20 μm, 2.3 μm, and 0.4 μm for Zn-0.8Ag, Zn-4.0Ag and Zn-4.0Ag-0.6Mn alloys, respectively [16,17], as consistent also with many other results [54]. Hence, the same phenomenon is expected to determine the hardness distribution in the alloys used in this investigation [16,28]. Additionally, the S2 alloy exhibits higher hardness in the HPT-CC state than the HPT-RS state and this may be due to these phenomena.

**4.2. Effect of rapid solidification combined with HPT on the thermal stability**

The ultrafine-grained microstructures produced by HPT in the alloys used in this investigation are significantly smaller than those generally reported for other materials. Based on earlier studies, it was evident that an evaluation of thermal stability is required to assess more fully the advantage of the current applied processing routes. The grain size of ~350 nm in the present investigation is significantly smaller than observed elsewhere and therefore heat treatment was performed from a relatively low temperature of 140 °C to a temperature of 220 °C for 10 minutes. Fig. 9 depicts the impact of varying heat treatment temperatures on the microstructure and hardness of the HPT-RS alloys. The primary focus of this research lay with the rapid solidification and thus changes in the HPT-RS samples were examined following heat treatment.

For the S1 alloy there was a minor grain growth even at 140 °C and this growth became more pronounced with increasing temperature. The presence of two intermetallic phases, $LiZn_4$ and $MnZn_{13}$, in the S1 alloy was expected to inhibit grain growth. Additionally, the HPT-RS samples contained a significant number of oxide particles that could act as obstacles. Fig. 9 provides evidence of simultaneous sluggish growth in both the $LiZn_4$ and Zn grains, suggesting that $LiZn_4$ is an inefficient phase for stabilizing ultrafine grains. The $MnZn_{13}$ phase was not observed in the HPT-RS microstructure but based on the phase diagram, it is expected to occur in the microstructure in the form of nanometric precipitates. The $MnZn_{13}$ phase effectively hindered grain growth at room temperature. However, according to the Mn-Zn phase diagram, heat treatment may dissolve $MnZn_{13}$ particles at elevated temperatures thereby negating their positive effect. Oxide particles in the HPT-RS samples impede grain boundary movement in the S1 HPT-RS samples but their quantity is not sufficient to effectively retain a nanometric grain size at elevated temperatures.

Similar results were observed in the S2 alloy with minor additions of Mg and Cu but a lower Mn content. The only notable difference is the presence of a small amount of $Mg_2Zn_{11}$ phase in the S2 alloy which acts similarly to $MnZn_{13}$ at room temperature and remains effective at high temperatures due to its limited solubility in Zn at elevated temperatures. A direct comparison of microstructure stability between the S1 and S2 alloys is challenging due to the minor but significant differences in chemical composition. Both alloys behave similarly up to 190 °C, while at 220 °C grain growth remains uniform in S1 and becomes heterogeneous in the S2 alloy. The hardness results show that heat treatment at all temperatures has no effect on the hardness in the S1 alloy, while in S2 there is a significant drop in hardness at 190 °C possibly due to the partial dissolution of the $CuZn_5$ and $Mg_2Zn_{11}$ phases. An $MnZn_{13}$ phase dissolution probably makes no contribution to the hardness drop since the phase was not visible in the S1 alloy.

Alloys heat-treated at 190 °C for 10 minutes were chosen for further mechanical evaluation due to the notable microstructural and hardness changes in the HPT-RS S1 and S2 alloys. Additionally, a direct comparison of microstructure was performed between the HPT-CC-HT and HPT-RS-HT samples to investigate the effect of rapid solidification on thermal stability. Fig. 10 illustrates the beneficial effect of rapid solidification on the grain size stability. Both HPT-RS-HT alloys exhibit grain sizes of about 800 nm while the grain size in the HPT-CC-HT alloys reaches 2 μm. Apart from grain size, significant differences are observed in the presence of coarse Cu- and Mg-rich particles in the HPT-CC S2 alloy which do not impede grain growth. The major hindering of grain growth is therefore attributed to the refined Mn-, Mg- and Cu-rich particles and oxides introduced during processing. Similarly, ZnO has successfully stabilized ultrafine-grained Zn produced by powder metallurgy and hot extrusion [55].

### 4.3. Evaluation of the complex microstructure-mechanical properties relationship in multicomponent Zn alloys

Grain refinement and microstructure stabilization are crucial factors that determine the mechanical properties required for biodegradable metallic materials. Evaluating the mechanical properties is of major importance and the tensile properties of the investigated alloys, as shown in Fig. 11, depend significantly on the processing route and chemical composition.

The HPT-CC S1 alloy exhibits relatively low strength but exceptional ductility compared to a similar alloy produced using conventional methods. For example, the UTS is equal to 395 MPa with an $E_f$ of 46% in Zn-0.48Li [13] while the UTS is 520 MPa with an $E_f$ of 6% in Zn-0.4Li or UTS of 450 MPa with an $E_f$ of 75% in Zn-0.4Li-0.4Mn [8]. The reduced mechanical properties of HPT-CC S1 are attributed to GBS which is activated by the severe grain refinement [5,36]. On the other hand, rapid solidification effectively enhances the strength of the S1 alloy while maintaining the ductility. HPT-RS S1 shows a 25% higher UTS than HPT-CC S1 due to the uniformly dispersed phases and oxides which generate a GBS pinning force [56].

In the HPT-CC S2 alloy, a small addition of Mg and Cu leads to a 30% increase in strength. Both Mg and Cu have been proven effective in enhancing the mechanical properties of Zn alloys. Thus, Mg forms nanometric precipitates within grains hindering dislocation movement [25,26,57] while Cu acts as a solid-solution strengthening element creating nanometric clusters and precipitates within the grains [58,59]. However, despite these positive alloying effects, GBS continues to play a major role in controlling the deformation behavior in the S2 alloy. Rapid solidification is less effective in this alloy providing only an 8% increase in strength while preserving the ductility.

The fabrication technique not only affects the mechanical properties but also influences the deformation behavior. The SRS measurements, performed using nanoindentation strain rate jump tests, reveal the impact of the processing route and heat treatment on the deformation mechanisms governing the plasticity of these alloys. The SRS data presented in Fig. 12 were used to calculate the activation volume $V^*$ using the following equation:

$$V^* = \sqrt{3}kT \frac{\partial \ln \dot{\varepsilon}}{\partial \sigma} = C\sqrt{3}kT \frac{\ln \dot{\varepsilon}_2 - \ln \dot{\varepsilon}_1}{H_2 - H_1} \qquad (2)$$

where $C$ is the Tabor factor equal to 3 [60], $k$ is Boltzmann's constant and $T$ is the temperature [40,61,62]. The calculated $V^*$ values were normalized to the Burgers vector of Zn ($\boldsymbol{b}$ = 2.66 x 10$^{-10}$ m). The measured $m$ and $V^*$ values for HPT-CC S1 alloy are approximately 0.18 and 6.7$\boldsymbol{b}^3$, respectively. These values are compared with typical values for rate-limiting processes: ~$\boldsymbol{b}^3$ for lattice or grain boundary diffusion, about 3-10$\boldsymbol{b}^3$ for GBS and other thermally activated processes and approximately 100-1000$\boldsymbol{b}^3$ for dislocation glide [63,64]. The comparison indicates that GBS is the dominant deformation behavior in the HPT-CC S1 alloy. The presence of dispersed second-phase precipitates and oxide particles acts as GBS pinning sources, reducing the value of $m$ to 0.14 and slightly increasing $V^*$ to 8.6$\boldsymbol{b}^3$. However, both values are consistent with a relatively high contribution of GBS to the total deformation. Furthermore, the addition of Mg and Cu, together with the associated grain refinement, promotes strain rate-dependent deformation mechanisms. The SRS and activation volume are 0.20 and 5.0$b^3$ for the HPT-CC alloy and 0.17 and 6.2$b^3$ for the HPT-RS alloy, respectively. These values are lower than for the S1 alloy. Despite the higher numbers of pinning particles, grain refinement promotes GBS which counteracts the pinning effect.

Heat treatment was employed to address the known drawbacks of ultrafine-grained Zn alloys by increasing strength at the expense of plasticity [10,17]. Short-time annealing significantly impacted the mechanical properties of the HPT-CC alloys (Fig. 11). Grain growth was observed leading to a remarkable 30% and 40% increase in strength for the HPT-CC S1 and HPT-CC S2 alloys, respectively. Consequently, S2 HPT-CC-HT achieved impressive YS, UTS and $E_f$ values of 440 MPa, 490 MPa and 23%, respectively. The measured values place the S2 HPT-CC-HT alloy between the strongest and most ductile Zn alloys reported in the literature [8,15,20].

By contrast, the higher thermal stability of the HPT-RS alloys resulted in lesser changes in the mechanical properties. The presence of oxides, nanometric second phases and inhibited grain growth led to a marginal increase in mechanical properties in both HPT-RS-HT alloys while reducing ductility by a few percent. The heat treatment of the HPT-RS S1 alloy exhibited high microstructural stability, as was evident by a stable strain rate sensitivity $m$ of 0.10 in the intermediate strain rate range, regardless of the annealing temperature. The reduction in $m$ value also slightly decreased the activation volume to 7.5$b^3$. In the HPT-RS-HT S2 alloy, a less pronounced effect was observed, where heat treatment significantly reduced the strain rate sensitivity only at the highest strain rate. However, the activation volume remained low, between 4.5$b^3$-5.6$b^3$, despite the considerable grain growth. Although short-time annealing effectively reduced the strain rate sensitivity, rate-dependent deformation mechanisms such as GBS or diffusion remained active such that the planned effect of the heat treatment was only partially achieved.

The present results align with other reports of GBS and activation volume calculations in ultrafine-grained Zn alloys [63]. Additionally, an activation volume below 10$b^3$ for any grain size suggests a contribution from grain boundary or lattice diffusion in the deformation mechanisms. The relatively high deformation temperature (0.43 of the absolute melting temperature), high grain boundary density and high mobility of the Li atoms may activate the GBS process controlled by grain boundary diffusion [65–67]. However, it is important to note that a single deformation mechanism cannot account for the entirety of the deformation behavior. In the case of the Zn-0.5Cu alloy, which possesses a grain size of approximately 2 µm, GBS has been found to contribute to 45% of the total deformation at room temperature when subjected to deformation at the high strain rate of $10^{-2}$ $s^{-1}$ [68]. Therefore, the S1 and S2 alloys with grain sizes several times smaller are expected to be even more susceptible to thermally-activated processes. Furthermore, Zn exhibits a notable tendency for dynamic recovery and recrystallization at room temperature [48,69]. The nucleation of partial dislocations followed by subsequent annihilation is another mechanism often considered responsible for the low activation volume observed in nanocrystalline alloys [64]. In summary, the high SRS and low activation volumes observed in the HPT-processed samples can be attributed to a combination of factors, including GBS, grain boundary and lattice diffusion, dynamic dislocation nucleation and annihilation, dislocation slip, and dynamic recrystallization [64,67,68].

Combining the present results and published data, it is readily apparent that achieving simultaneous optimizations of high mechanical properties, good plasticity and high thermal stability poses a significant challenge. Designing load-bearing Zn alloys involves addressing several drawbacks. The most promising area for improving Zn alloys lies in effectively utilizing alloying elements, where a complex chemical composition enhances the mechanical properties when each alloying element serves a specific role. However, this approach is not sufficient to prevent GBS and in practice it is crucial to attain a uniform distribution of strengthening second-phase particles to achieve the desired GBS suppression. Nonetheless, this can also lead to grain refinement and thereby inadvertently promoting GBS. Therefore, having uniformly dispersed second phases acting as GBS pinning forces makes it easier to control other factors, such as the grain size.

This research provides a clear demonstration that rapid solidification enhances thermal stability and slightly reduces GBS in HPT-processed Zn alloys. While the same phenomena should be universal for other processing techniques, a comprehensive design must consider the proper chemical composition with solid-solution and precipitation-strengthening elements, creating GBS pinning particles and ensuring a uniform distribution of all microstructural constituents. Furthermore, grain size adjustment must effectively balance the beneficial H-P and detrimental GBS effects.

## 5. Conclusions

This study explores the influence of rapid solidification followed by high-pressure torsion on the microstructure, mechanical properties and microstructure stability of hypoeutectic Zn-0.33Li-0.39Mn and Zn-0.33Li-0.27Mn-0.14Mg-0.10Cu alloys. The investigation yields several important results:

- Rapid solidification has proven instrumental in achieving an unparalleled uniform distribution of alloying elements, a feat unattainable through conventional casting methods. This effect is particularly pronounced in the case of Mg-rich precipitates, with their lack observed in the rapidly solidified S2 alloy.

    Bulk Zn alloys with an ultrafine-grained structure were successfully fabricated by applying high-pressure torsion involving 15 rotations at 6 GPa pressure on melt-spun ribbons. Although exhibiting slight contamination by oxide impurities, these alloys achieved an average grain size below 360 nm.

- Heat treatment conducted at temperatures ranging from 140 °C to 220 °C for 10 minutes revealed markedly superior thermal stability in rapidly solidified alloys compared to their conventionally cast counterparts. Subsequent annealing at 190°C led to grain sizes of approximately 0.8 μm and 2.0 μm in rapidly solidified and conventionally cast alloys, respectively.

- Remarkably, the HPT-CC-HT S2 alloy exhibited exceptional mechanical properties, with a YS of 440±11 MPa, UTS of 490±5 MPa and an $E_f$ of 20%. This places the alloy in direct competition with high-Li Zn-Li-Mn variants.

- Examining the strain rate sensitivity and activation volume values indicates that grain boundary sliding and diffusion mechanisms primarily govern deformation in these ultrafine-grained alloys. Notably, while heat treatment and chemical composition exert an influence on the strain rate sensitivity, the activation volume remains essentially unaffected.

In summary, this study demonstrates that rapid solidification and high-pressure torsion can greatly enhance the microstructure and mechanical characteristics of Zn-based alloys. These enhanced attributes create opportunities for innovative uses, prompting continued exploration in the realm of advanced materials design.

**Data availability**

The data presented in this study are available in the repository of the corresponding author.

**Author contributions**

**W. Bednarczyk:** Conceptualization, Methodology, Investigation, Formal analysis, Visualization, Writing – Original draft, Resources, Funding acquisition, Project administration; **M. Wątroba**: Methodology, Investigation, Formal analysis, Writing - Review & Editing; **G. Cieślak**: Investigation; **M. Ciemiorek:** Investigation, Formal analysis; **K. Hamułka:** Investigation; **C. Schreiner:** Investigation, Formal analysis, Resources; **R. Figi:** Resources; **M. Marciszko-Wiąckowska:** Investigation; **G. Cios:** Investigation, Formal analysis; **J. Schwiedrzik:** Resources, Writing - Review & Editing; **J. Michler:** Resources; **N. Gao:** Resources, Writing - Review & Editing; **M. Lewandowska:** Resources, Writing - Review & Editing, Supervision; **T.G. Langdon**: Resources, Writing - Review & Editing, Supervision.

**Acknowledgments**

This research was funded in whole or in part by National Science Centre, Poland, under Grant UMO-2021/40/C/ST5/00071. For the purpose of Open Access, the author has applied a CC-BY public copyright licence to any Author Accepted Manuscript (AAM) version arising from this submission. WB was partially supported by the Foundation for Polish Science (FNP) with scholarship START 2023 (no. START 003.2023). TGL was supported by the European Research Council under ERC Grant Agreement No. 267464-SPDMETALS. Strain rate jump tests were performed with the financial support of the IDUB PW Mobility PW V program. The authors would like to sincerely thank Dr. Renato Pero and Alemnis AG for their technical, substantive and software help.